\documentclass[preprint,12pt]{elsarticle}

\usepackage{amssymb}

\begin{document}

\begin{frontmatter}

\title{Artifact Dark Matter from Unified Brane Gravity}

\author{Ilya Gurwich}
\ead{gurwichphys@gmail.com}
\author{Aharon Davidson}
\ead{davidson@bgu.ac.il}
\address{Physics Department, Ben-Gurion University,
Beer-Sheva 84105, Israel}

\begin{abstract}
Adopting Dirac's brane variation prescription,
the energy-momentum tensor of a  brane gets supplemented by a 
geometrical (embedding originated) dark component. While the
masslessness of  the graviton is preserved, and the Newton
force law is recovered, the corresponding Newton constant is
necessarily lower than the one which governs  FRW cosmology.
This has the potential to puzzle out cosmological dark matter,
a subsequent conjecture concerning  galactic dark matter follows.
\end{abstract}



\end{frontmatter}

\section{Introduction}

Recently the idea that brane theories \cite{brane} could accommodate the
dark matter phenomenon was suggested \cite{DMbranes}.
Brane theories, have recently made great breakthroughs in the area of
reproducing some results of general relativity, on the cosmic scale as
well as the regular Newton potential \cite{LBG}. Since brane theories
originated to solve the puzzle of quantum gravity by allowing the
existence of extra-dimensions, the possibility that they can naturally
produce a solution to a seemingly unrelated problem in gravity will
generate a great boost in the theory asides from being a significant
achievement and a good verification of the branes and extra-dimensions
ideas.

Despite the recent progress, there is no natural theoretical framework
for dark matter. We will try the approach of unified
brane gravity \cite{UBG}, following Dirac's prescription of careful
variation in the region of the brane \cite{Dirac}.

\section{General Perturbations and the Graviton }

We begin with the simplest scenario of a positive tension 4-dimensional
flat brane
embedded in 5-dimensional AdS bulk
\begin{equation}
	\displaystyle{ds^{2}_{5}=dy^2+e^{-2b\left|y\right|}
	\eta_{\mu\nu}dx^{\mu}dx^{\nu}} ~.
\end{equation}
$b^{-1}=\sqrt{-6/\Lambda_5}$ denotes the
AdS scale, $\eta_{\mu\nu}$ is the 4-dimensional Minkowski
metric, and the brane is conveniently located at $y=0$.
Before turning to the main discussion concerning perturbations
of this brane, it is imperative to understand the full potential of
the unperturbed brane.
In the conventional RS (Randall-Sundrum), DGP (Dvali-Gabadadze-Porrati)
and CH (Collins-Holdom)  scenarios, in order to ensure
its flatness, the brane has to be of positive (or negative) tension
\begin{equation}
	\sigma=\frac{3b}{4\pi G_5} ~.
	\label{tens}
\end{equation}
Unified brane gravity (UBG), although requires the same, allows
for one more degree of freedom.
To see the point, first recall that the UBG field equations are given by
\begin{equation}
	\displaystyle{\frac{1}{4\pi G_5}
	\left(K_{\mu\nu}-g_{\mu\nu}K\right)=
	\frac{3b}{4\pi G_5}g_{\mu\nu}
	+\frac{1}{8\pi G_4}
	\left(R_{\mu\nu}-\frac{1}{2}
	g_{\mu\nu}R\right)+
	T_{\mu\nu}+\lambda_{\mu\nu}} ~.
	\label{FieldEq}
\end{equation}
In addition to the familiar terms (namely, the Israel junction term,
the brane surface tension, the Einstein tensor associated with the
scalar curvature $R_4$, and the physical energy-momentum
tensor $T_{\mu\nu}=\delta {\cal L}_{matter}/\delta g^{\mu\nu}$
of the brane), UBG introduces $\lambda_{\mu\nu}$.
The latter consists of Lagrange multipliers associated with the
fundamental induced metric constraint
$g_{\mu\nu}(x)=g_{MN}(y(x))y^{M}_{,\mu}y^{N}_{,\nu}$.
In the above field equations, $\lambda_{\mu\nu}$ serves as a geometrical
(embedding originated) contribution to the total energy-momentum
tensor of the brane.
$\lambda_{\mu\nu}$ is furthermore conserved, and its contraction with the
extrinsic curvature vanishes
\begin{equation}
	\lambda^{\mu\nu}_{~~;\nu}=0~, \quad
	\lambda_{\mu\nu}K^{\mu\nu}=0 ~.
	\label{ubgbasic}
\end{equation}
By choosing $\lambda_{\mu\nu}=0$, which is a viable choice,
one approaches the conventional DGP(CH) limit.
For a flat brane embedded in a 5-dimensional AdS background,
which is the special case of interest, $K_{\mu\nu}=-b\eta_{\mu\nu}$.
In turn, eq.(\ref{ubgbasic}) simply implies that the corresponding
$\lambda_{\mu\nu}$ is traceless.
A traceless and conserved source serves as an effective (positive or
negative) radiation term.
The flatness of the unperturbed brane can be achieved, the conventional
way, if the energy-momentum and the embedding terms both vanish,
that is $T_{\mu\nu}=\lambda_{\mu\nu}=0$.
But now there exists the milder option $T_{\mu\nu}+\lambda_{\mu\nu}=0$.
Following the above, if (and only if) the real matter on the brane exclusively
consists of radiation, one can choose an appropriate $\lambda_{\mu\nu}$
to cancel it out.
To be more specific, let our unperturbed flat brane host a constant radiation
density $\rho$, and choose the embedding counter term to be
$\lambda^{0}_{\mu\nu}=-T^{0,rad}_{\mu\nu}=
-diag\left(\rho,\frac{1}{3}\rho, \frac{1}{3}\rho, \frac{1}{3}\rho\right)$.
Reflecting the peculiarity that \emph{a flat brane can in fact be hot},
which is unique to UBG, the perturbations are expected to be quite
different from those around a DGP brane, thus giving rise to new physics.
Since for a general perturbation, $\delta K^{\mu\nu}$ is not
proportional to $h^{\mu\nu}$, the term
\begin{equation}
	s_{\mu\nu}\equiv \lambda_{\mu\nu}+T^{rad}_{\mu\nu}
	=\delta\lambda_{\mu\nu}+\delta T^{rad}_{\mu\nu}
\end{equation}
is not necessarily zero.
One can furthermore verify that $s_{\mu\nu}$ is conserved, and not necessarily
traceless
\begin{equation}
	s\equiv \eta^{\mu\nu} s_{\mu\nu}=
	\frac{1}{2b}\lambda^{0}_{\mu\nu}\left(\frac{\partial}{\partial\left|y\right|}
	+2b\right)h^{\mu\nu} ~.
	\label{trace}
\end{equation}
The non-localized part of the perturbation equations is the same as the
familiar RS case, since the bulk still follows the normal
5-dimensional Einstein equations
\begin{equation}
	\displaystyle{\left(\frac{\partial^2}{\partial\left|y\right|^2}
	-4b^2+e^{2b\left|y\right|}
	\fbox{\scriptsize 4}\right)
	h_{\mu\nu}=0,}
	\label{5dpert}
\end{equation}
where $\fbox{\scriptsize 4}\equiv
\eta^{\mu\nu}\partial_{\mu}\partial_{\nu}$
is the 4-dimensional (unperturbed) d'Alembertian.
The localized part of the equation is
\begin{equation}
	\displaystyle{\delta(y)\left[\frac{1}{8\pi G_5}
	\left(\frac{\partial}{\partial\left|y\right|}
	+2b\right)+\frac{1}{8\pi G_4}
	\fbox{\scriptsize 4}\right]h_{\mu\nu}
	=\delta(y)\left(\tau_{\mu\nu}+
	s_{\mu\nu}\right).}
	\label{4dpert}
\end{equation}
The propagation of modes into the bulk remains the same as in all the familiar
cases. Thus, we will only be focusing on the perturbations on the brane.
expanding the solution into bulk mass modes,
$h_{\mu\nu}=A(y)\bar{h}_{\mu\nu}\left(x^{\mu}\right)$,
where we normalize without loss of generality $A(0)=1$ and define
$\displaystyle{\alpha=1+\frac{1}{2b}A^{\prime}(0)}$.
Next let us separate the perturbation,
$\bar{h}_{\mu\nu}=h^{(m)}_{\mu\nu}+h^{(u)}_{\mu\nu}$,
to the standard term $h^{(m)}_{\mu\nu}$,
which follows the usual brane equation and thus admits the familiar solutions
and the new term $h^{(u)}_{\mu\nu}$, which is a direct result of the
additional effective source $s_{\mu\nu}$.
Unfortunately we cannot find a general Green function to 
eq.(\ref{4dpert},\ref{5dpert}),
because there is no closed form to express $s_{\mu\nu}$ in term of
$h^{(u)}_{\mu\nu}$. To that end, the only general prescription to
solve these equations is perturbatively in $\rho$.
when expanding the cosmological equations around a flat
background with positive tension and radiation density $\rho$,
we get the FRW equation for the brane,
\begin{equation}
	\displaystyle{\delta\rho =
	\left(\frac{1}{8\pi G_{4}}+\sqrt{\frac{6}{-\Lambda_{5}}}
	\left(\frac{1}{8\pi G_{5}}+
	\frac{\rho}{6b}\right)
	\right)\epsilon~,}
\end{equation}
where $\displaystyle{\epsilon=3\frac{\dot{a}^{2}+k}{a^{2}}}$
and therefore, the corresponding Newton constant is,
\begin{equation}
	\displaystyle{\frac{1}{G^{c}_N}=\frac{1}{G_{CH}}+
	\frac{4\pi\rho}{3b^2}.}
	\label{GN6}
\end{equation}
Where the $c$ stands for cosmological and $G_{CH}$ is the Newton
constant that corresponds to the CH scenario,
\begin{equation}
	\displaystyle{\frac{1}{G_{CH}}=\frac{1}{G_{4}}+
	\frac{1}{G_{RS}},}
	\label{GNCH}
\end{equation}
where $G_{RS}=bG_5$.

\section{Static Radial Source and Dark Matter}

For the radial case we can write the equation for $h^{(u)}$
\begin{equation}
	\displaystyle{\kappa^2_4rh^{(u)\prime\prime\prime}+
	4\kappa^2_4h^{(u)\prime\prime}+
	\left(\frac{2\kappa^2_4}{r}+\left(k-\frac{2}{3}
	\alpha\rho\right)r\right)
	h^{(u)\prime}
  +2kh^{(u)}=
	-\frac{4G_{CH}M\alpha\rho}{3r},}
	\label{radpert}
\end{equation}
where $M$ is the mass of the physical source,
$\displaystyle{\kappa^2_4\equiv\frac{3}{16\pi G_4}}$ and
$\displaystyle{k\equiv\frac{\alpha b^2}{2\pi G_{RS}}}$. 
The solution for the full perturbation yields
$\bar{h}_{\mu\nu}=h^{(m)}_{\mu\nu}+h^{(u)}_{\mu\nu}$ is therefore
\begin{equation}
	\bar{h}_{tt}=\bar{h}_{rr}=
	\frac{1}{\displaystyle{1+\frac{4\pi G_{RS}\rho}{3b^2}}}\frac{2G_{CH}M}{r},
\end{equation}
It is important to note that it is only due to the solution
being independent of $\alpha$ that we can proceed without integrating
over all the mass modes.
The Newton potential is thus recovered, giving us further
reassuring that the graviton is indeed massless, since a mass term
in the propagator would have to have generated an exponential decay.
The Newton constant associated with the solution is
\begin{equation}
	G^{r}_N=\frac{G_{CH}}{\displaystyle{1+\frac{4\pi G_{RS}\rho}{3b^2}}}
	\label{GN1}
\end{equation}
where the $r$ index stands for radial.

\smallskip

Now that the mathematics has been understood, we return to physics.
Alone, eq.(\ref{GN1}) has nothing new to offer.
However, if we compare the cosmological and radial result, we see
that the Newton constants differ and we need to see, how significant
is this difference.
First of all since we do have bounds on $b$ both from particle
and gravitational localization, we can clearly state that the term
$\frac{\rho}{b^2}$ is negligible in both equations.
This means that $G^{c}_{N}=G_{CH}$, whereas,
\begin{equation}
  \displaystyle{\frac{1}{G^{r}_N}=\frac{1}{G^{c}_N}+
	\frac{4\pi\rho}{3b^2}\frac{G_{RS}}{G_4}.}
	\label{GN4}
\end{equation}
The last term in the radial gravitational constant would have been
negligible if not for the factor
$\displaystyle{\frac{G_{RS}}{G_4}}$. We have no experimental or theoretical bounds on
the latter ratio. In fact the proposed self accelerated DGP solution for the
cosmological constant, requires this quantity to be very large. If it is
large enough, then this term can be significant in the calculation of the
Newton constant. Thus, in principle we have a real difference between the
cosmological and the radial gravitational constant. The radial constant
being necessarily lower. However, historically, the Newton constant was measured in
radial systems (solar system). And thus \emph{an observer that is unfamiliar
with this physics, would interpret this effective growth of the
gravitational constant as missing cosmological mass} (since in general
relativity, mass is inseparable from the gravitational constant). Thus
bringing him to the phenomenon of cosmological dark matter, \emph{without
facing dark matter in the solar system}.

When solving the perturbation equation in cosmic background, one expects the
two branches of the solution, one being the $G^{r}_{N}$ and the other
$G^{c}_{N}$ to be connected, creating some sort of transition between them.
Such transition, to an unaware observer, will seem as a gradual increase of mass,
that may result in flat rotation curves (FRC).
Although the exact solution to fluctuations around a cosmological brane is
highly complex. We can give a rough estimate to the typical scale of such a
transition, thus formulating our conjecture. We assume the scale to be
roughly in the region where cosmological and radial curvatures are of the
same order of magnitude,
so that cosmology and radial solutions \emph{"mix"}. The radial curvature
is of the order $\displaystyle{\frac{r_s}{r^3}}$, $r_s$ being the Schwarzschild radius
and the cosmological is of the order of $H^2$, $H$ being the Hubble constant.
The scale of the predicted FRC is therefore
\begin{equation}
	\displaystyle{r_{FRC}\sim\left(r_st^{2}_{Hubble}\right)^{1/3}\propto M^{1/3},}
	\label{FRC}
\end{equation}
where $t_{Hubble}$ is the age of the universe. When this scale is calculated for the
sun, the result is 100 ly, which is way beyond the scale of the solar
system. At these distances, other stars contribute and thus the effect is
unmeasurable today. For a galactic mass on the other hand the result is of the order
of $10^5$ light years, which is only one order of magnitude higher than
the real galactic scale. One needs to remember that it is only a rough estimate
and also that galaxies are composed of many stars, each giving an effect on the
scale of about 100 ly, so that the combined effect may be closer that the above
result, to give the exact scale of FRC.

\end{document}